\documentclass[multphys,vecphys]{svmult}

\usepackage{makeidx}         
\usepackage{graphicx}        
\usepackage{multicol}        
\usepackage[bottom]{footmisc}

\makeindex             

\begin{document}

\title*{A Wide-Field Survey of the Globular Cluster Systems of Giant Galaxies}
\author{Katherine L.\ Rhode}
\institute{Astronomy Department, Wesleyan University, Middletown, CT 06459, USA; kathy@astro.wesleyan.edu}
%
%
\maketitle

\subparagraph{Abstract.} I present selected results from a wide-field CCD survey of the GC
systems of giant galaxies, including showing how measurements of the
specific frequency of metal-poor GCs can constrain the redshift of
their formation.


\section{Introduction to the Survey}
\label{section:intro}


\subsection{Motivation and Design}
\label{section:motivation}
Measurements of the ensemble properties of the globular cluster (GC)
systems of massive galaxies can provide a direct test of theories for
how such galaxies form.  Making these global measurements requires
observations of all or most of a galaxy's GC system, which in turn
requires {\bf wide-field imaging}.  In recent years Steve Zepf
(Michigan State) and I have been collaborating on a wide-field CCD
survey of the GC systems of giant galaxies.  Some motivations for the
survey are that: the total numbers of GCs in giant galaxies are
uncertain; few spiral galaxies have been studied; and the outer
regions of GC systems are largely unexplored, at least by modern CCD
studies.

To date we have $BVR$ or $BVI$ imaging of four E/S0 galaxies from the
Kitt Peak 4m telescope and Mosaic Imager, and nine spiral galaxies
from large-format CCDs on the WIYN 3.5m telescope.  At the distances
of the targets (10$-$20~Mpc), the GCs are unresolved so we use
three-color photometry and good image resolution to select them and to
exclude stars and background galaxies.  Our aim is to derive reliable
global properties of the GC systems and to use them to study galaxy
formation and test models for the origin of ellipticals, such as
spiral mergers (Ashman \& Zepf 1992), multi-phase collapse (Forbes et
al.\ 1997), and collapse with accretion (C\^ot\'e et al.\ 1998).

\subsection{Data and Analysis Methods}
\label{section:methods}
The spiral galaxy images cover 7' x 7' or 10' x 10', depending on the
WIYN detector used.  This translates to radial coverage of 30$-$40
kpc.  Early-type galaxies often have extended GC systems, so the
images of the E/S0s are 38' x 38' and provide radial coverage of
60$-$120~kpc.  To find GCs, we create a deep, stacked image in each
filter; remove the diffuse galaxy light; find sources above a chosen
S/N level; discard extended objects; and select point sources with
$BVR$ or $BVI$ magnitudes and colors like GCs.  We also run
completeness tests and fit the GCLF to determine what fraction of GCs
are missing given our detection limits. We use Galactic star count
models and archival HST data to quantify the contamination that
remains from stars and galaxies.

\section{Selected Results}
\label{section:results}
The survey has so far yielded positions and $BVR$ photometry of
hundreds to thousands of GC candidates in eight galaxies.  In all
cases we have imaged the full radial extent of the systems.  The
derived properties --- e.g., number of GCs, spatial and color
distributions, color gradient --- are thus global ones and provide an
important comparison to model predictions.  We find that all the
models mentioned in Sect.~\ref{section:motivation} have
inconsistencies with the data (see Rhode \& Zepf 2004).  Below I
highlight two general results from the survey.

\subsection{Total Numbers \& Specific Frequencies}
\label{section:S_N}
We calculate the total number of GCs ($N_{GC}$) in each galaxy by
integrating the derived radial distribution of the GC system from the
galaxy center to the radius at which the GC surface density goes to
zero within the errors.  Table~\ref{table:S_N} gives $N_{GC}$ and
specific frequency ($S_N$ from Harris \& van den Bergh 1981) for the
eight galaxies analyzed.  Columns (1)$-$(3) are galaxy name, type, and
magnitude, columns (4)$-$(5) are $N_{GC}$ and $S_N$ from the survey,
and columns (6)$-$(7) list, when applicable, $S_N$ from past work and
a reference.  ($S_N$ from past work combines $N_{GC}$ from the study
with the $M_V$ we assumed.)  Our $S_N$ values are 20$-$75\% lower for
four of the six galaxies in Table~\ref{table:S_N} that were studied
previously, and our errors on $S_N$ are 2$-$4 times smaller for all
six galaxies.  The smaller values are due at least in part to reduced
contamination levels.  Also, observing the full extent of the GC
systems yields smaller, better-constrained total numbers than
observing the inner GC system and extrapolating to an arbitrary outer
radius, as is typical in past CCD work.

\begin{table}
\centering
\caption{Total number and $S_N$ for eight survey galaxies}
\label{table:S_N}
\begin{tabular}{lcrcrcccr}
\hline\noalign{\smallskip}
Galaxy & Type & $M_V$ & & $N_{GC}$ & & $S_N$ & Previous $S_N$ & Reference \\
\noalign{\smallskip}\hline\noalign{\smallskip}
NGC~4472 & E2 & $-$23.1 & & 5900 & & 3.6$\pm$0.6 & 4.5$\pm$1.3 & Harris 1991\\
NGC~4406 & E3 & $-$22.3 & & 2900 & & 3.5$\pm$0.5 & 4.6$\pm$1.1 & Harris 1991\\
NGC~4594 & S0 & $-$22.4 & & 1900 & & 2.1$\pm$0.3 & 2$\pm$1   & Bridges \&
Hanes 1992\\
NGC~3379 & E1 & $-$20.9 & & 270  & & 1.2$\pm$0.3 & 1.1$\pm$0.6 & Harris 1991\\
NGC~7814 & Sab& $-$20.4 & & 170  & & 1.3$\pm$0.4 & 5.2$\pm$1.7 & Bothun
et al.\ 1992\\
NGC~3556 & Sc & $-$21.2 & & 290  & & 0.9$\pm$0.4 & ... & ...\\
NGC~2683 & Sb & $-$20.5 & & 120  & & 0.8$\pm$0.4 & 2.0$\pm$0.7 & Harris et
al.\ 1985\\
NGC~4157 & Sb & $-$20.4 & & 80   & & 0.6$\pm$0.3 & ... & ...\\
\noalign{\smallskip}\hline
\end{tabular}
\end{table}

\subsection{Mass-Normalized Number of Blue (Metal-Poor) GCs}
\label{section:Tblue}
\begin{figure}
\centering
\includegraphics[height=6cm]{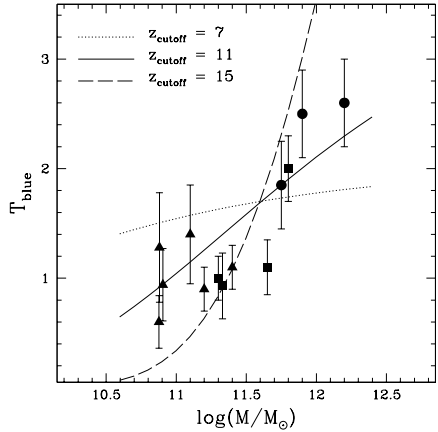}
%
%
\caption{$T_{\rm blue}$ vs.\ log of the galaxy mass for 8 survey galaxies
  and 5 from the literature.  Circles are cluster elliptical galaxies,
  squares are field E/S0s, and triangles are field spirals. The curves
  show the expected trend if metal-poor GCs form prior to $z$ of 7,
  11, or 15 (details in Sect.~\ref{section:Tblue}).}
\label{fig:Tblue}       
\end{figure}

Ashman \& Zepf (1992) proposed that elliptical galaxies form in spiral
galaxy mergers and that the metal-poor GCs in ellipticals come
directly from the progenitor spirals.  If this is true, the
galaxy-mass-normalized number of metal-poor GCs (called $T_{\rm
blue}$) should be similar for spiral and elliptical galaxies.  The
well-determined global values from the survey allow us to test this
prediction. Figure~\ref{fig:Tblue} shows $T_{\rm blue}$ vs.\ log of
the stellar mass for eight survey galaxies, plus the Milky Way, M31,
and three galaxies from the literature (see Rhode, Zepf, \& Santos
2005 for a list).  $T_{\rm blue}$ is defined as ${N_{GC}(\rm
blue)}/{(M_G/10^9\ {\rm M_{\odot}})}$.  $T_{\rm blue}$ for the spiral
galaxies is too small to account for the blue GC populations of the
giant cluster ellipticals, implying that the merger model cannot
explain their formation.  The data also show a trend of increasing
$T_{\rm blue}$ with increasing galaxy mass.  This is generally
consistent with a galaxy and GC system formation scenario described by
Santos (2003) that combines biased structure formation with
hierarchical merging.  In this model, metal-poor GCs form over a
finite period in the early Universe, when gas-rich protogalactic
fragments merge into larger structures. Structure formation is
temporarily suppressed at high-$z$, perhaps due to cosmic
reionization.  Meanwhile, stellar evolution enriches the intergalactic
medium, so any GCs formed later are metal-rich compared to the first
generation of GCs.  Today's massive galaxies in high-density regions
of the Universe began assembling first, so formed relatively more
metal-poor GCs during the initial formation epoch.

%
%

Figure~\ref{fig:Tblue} shows the expected trend if the formation
cutoff for the first generation of GCs occurred at $z$ $=$ 7, 11, or
15.  The curves come from an extended Press-Schechter calculation by
G.\ Bryan (Columbia) that assumes that $T_{\rm blue}$ is proportional
to the fraction of a galaxy's mass that has collapsed into halos of at
least 10$^8$~M$_{\odot}$ by the truncation redshift.  Although a
biased, hierarchical scenario appears generally consistent with our
$T_{\rm blue}$ data, more modeling and data are needed to determine
whether it can reproduce all of the properties
of the galaxy GC systems we observe.

\section{Ongoing and Future Work}
\label{section:future work}
\subparagraph{Imaging.} We are analyzing WIYN data 
of more spiral galaxies, 
have begun using an 8Kx8K mosaic CCD on the MDM-2.4m telescope to
image more targets, and also plan to use the WIYN One Degree Imager,
which will be commissioned in 2009 and provides 0.5'' resolution over
a 1$^o$ field.

\subparagraph{Follow-Up Spectroscopy.} With our collaborators, we
are using telescopes like the AAT, VLT, Keck, and WIYN to obtain
spectra and derive radial velocities of the GC candidates, in order to
study the kinematics of the GC systems and measure the mass profiles
of the host galaxies to 10$-$15 $R_{\rm eff}$.  Our analysis of the
mass distribution of NGC~3379 is published in Bergond et al.\ 2006.

\subparagraph{New Model Predictions.} The ideas about blue GCs and
biasing in Sect.~\ref{section:Tblue} were introduced in Santos (2003)
and based on models of the formation of the Milky Way and its GC
system.  M.\ Santos (STScI) is modeling the formation of galaxies of
varying masses and environments and we are working with him to develop
meaningful comparisons between our data and the simulations.

\bigskip
\noindent The author is supported by an NSF
Astronomy \& Astrophysics Postdoctoral Fellowship under award
AST-0302095.


\end{document}